\documentclass[10pt, twocolumn, secnumarabic, amssymb, nobibnotes, footinbib, showpacs, aps, pra, superscriptaddress]{revtex4-1}

\usepackage{graphicx}
\usepackage{dcolumn}
\usepackage{amsmath}
\usepackage{natbib}
\usepackage{color}
\usepackage{epstopdf}
\usepackage{epsfig}

\def\identity{\leavevmode\hbox{\small1\kern-3.8pt\normalsize1}}

\newtheorem{propo}{Proposition}
\newcommand{\be}{\begin{eqnarray}}
\newcommand{\ee}{\end{eqnarray}}
\newcommand{\bpr}{\begin{propo}}
\newcommand{\epr}{\end{propo}}
\newcommand{\bpf}{\begin{proof}}
\newcommand{\epf}{\end{proof}}
\newcommand{\ket}[1]{\left | #1 \right\rangle}
\newcommand{\bra}[1]{\left \langle #1 \right |}

\newcommand{\Tr}{\mathrm{Tr}}

\newcommand{\proj}[1]{\ket{#1}\bra{#1}}

\renewcommand{\epsilon}{\varepsilon}
\begin{document}

%%%%%%%%%%%%%%%%%%%%%%%%%%%%%%%%%%%%%%%%%%%%%%%%%%%%%%%%%%%%%%%%%%%%

\title{Investigating nonclassicality of many qutrits by symmetric two-qubit operators}

\author{Marcin Markiewicz} \email{marcin.1.markiewicz@uj.edu.pl}    
\affiliation{Institute of Physics, Jagiellonian University, ul. \L{}ojasiewicza 11,
30-348 Krak\'ow, Poland}

\author{Kamil Kostrzewa}
\author{Adrian Ko{\l}odziejski}
\affiliation{Institute of Theoretical Physics and Astrophysics, Faculty of Mathematics, Physics and Informatics, University of Gda\'nsk, 80-308 Gda\'nsk, Poland}

\author{Pawe\l{} Kurzy\'nski}   \affiliation{Faculty of Physics, Adam Mickiewicz University, Umultowska 85, 61-614 Pozna\'n, Poland}
\author{Wies{\l}aw Laskowski}
\affiliation{Institute of Theoretical Physics and Astrophysics, Faculty of Mathematics, Physics and Informatics, University of Gda\'nsk, 80-308 Gda\'nsk, Poland}

\date{\today}

%%%%%%%%%%%%%%%%%%%%%%%%%%%%%%%%%%%%%%%%%%%%%%%%%%%%%%%%%%%%%%%%%%%%

\begin{abstract}
We introduce a new method of investigating qutrit nonclassicality by translating qutrit operators to symmetric two-qubit operators. We show that this procedure partially resolves the discrepancy between maximal qutrit entanglement and maximal nonclassicality of qutrit correlations. Namely we express Bell operators corresponding to qutrit Bell inequalities in terms of symmetric two-qubit operators, and analyze the maximal quantum violation of a given Bell inequality from the qubit perspective. As an example we show that the two-qutrit CGLMP(Collins-Gisin-Linden-Massar-Popescu) Bell inequality can be seen as a combination of Mermin's and CHSH (Clauser-Horne-Shimony-Holt) qubit Bell inequalities, and therefore the optimal state violating this combination differs from the one which corresponds to the maximally entangled state of two qutrits. In addition, we discuss the same problem for a three qutrit inequality. We also demonstrate that the maximal quantum violation of the CGLMP inequality follows from complementarity of correlations. 
\end{abstract}

\pacs{03.65.Ud}

\maketitle

%%%%%%%%%%%%%%%%%%%%%%%%%%%%%%%%%%%%%%%%%%%%%%%%%%

\section{Introduction}

Quantum entanglement and nonclassicality of quantum correlations seemed to appear as two sides of the same coin. However some discrepancies from this picture were found. Firstly, it was shown that there exist weakly entangled states that do not give rise to nonclassical correlations \cite{Werner89}. Further it turned out that two- and three-qutrit states reveal maximal nonclassicality for non-maximally entangled states \cite{CGLMPorigin, AcinChen04, LRZ14, Gruca12}. In the first case the discrepancy between nonclassicality of correlations and entanglement was reduced by showing, that any bipartite entangled state gives rise to nonclassical correlations, if properly extended by attaching a classically correlated state \cite{Masanes08}. In the second case the discrepancy has been questioned by suggesting that the maximal violation of an optimal Bell inequality is not a proper measure of maximal nonclassicality \cite{CGLMPvolume}. In this work we show that the last discrepancy disappears if we translate optimal quantum correlations of qutrits into a qubit representation. For this aim we introduce a new way of analyzing maximal quantum violation of Bell inequalities by many qutrit states. Namely, we represent the optimal qutrit measurement operators by means of symmetric two-qubit operators.

Nonclassical nature of quantum correlations in the case of two three-level systems (qutrits) was firstly demonstrated numerically by Kaszlikowski et. al \cite{Kaszlikowski00}. However the first analytical form of a Bell inequality for qutrit states was found a few years later by Collins et. al. In \cite{CGLMPorigin} they proposed a set of Bell inequalities (called further CGLMP inequalities) for bipartite correlations, with two settings per observer and arbitrary number $d$ of outcomes, which are violated by quantum $d$-level systems. A little bit later a paradoxical nature of two-qutrit nonclassicality has been revealed: although the CGLMP inequalities are optimal \cite{Masanes02}, they are maximally violated by  non-maximally entangled two-qutrit states \cite{CGLMPacin}. What is more, the discrepancy between the CGLMP violation for maximally and non-maximally entangled states increases with the system's dimension \cite{CGLMPacin, CGLMPmax08}. 
Further a similar effect was found in the case of three qutrits, namely Acin et. al. \cite{AcinChen04} found a generalization of a CGLMP inequality to a three-qutrit case, which is tight and maximally violated by a non-maximally entangled state.
Although this discrepancy between maximal nonclassicality and maximal entanglement has been thoroughly studied from the geometrical perspective \cite{Spengler11, CGLMPvolume}, it is still lacking a deeper understanding.

In this work we present a new approach to the analysis of a qutrit nonclassicality --- namely we analyse the form of the Bell operator \cite{Braunstein92} corresponding to the qutrit Bell inequalities in two different local operator bases, completely different from the ones used in \cite{CGLMPmax08}: the spin-$1$  basis in $3$-dimensional representation \cite{SpinSqueezed93} and the spin-1  basis in 4-dimensional representation, which corresponds to the symmetric subspace of two qubits \cite{Kurzynski16}. 
Expressing qutrit Bell operators in the local bases of symmetric-two-qubit operators allows for translating the analysis of a maximal qutrit nonclassicality to the analysis of many qubit nonclassicality --- the topic, which is much more understood and intuitive.
Using this method we show that the CGLMP Bell operator in the four-qubit symmetric subspace is a composition of correlations corresponding to CHSH \cite{CHSH} and Mermin's \cite{Mermin90} inequalities, therefore the optimal state for its violation is a superposition of states maximizing violations of CHSH and Mermin inequalities respectively, that is a superposition of two two-qubit Bell states and the four qubit GHZ state. The maximally entangled state of two qutrits in the four qubit representation is also of this form, however with slightly different superposition coefficients. 
Moreover using the four-qubit representation we show, that the maximal quantum violation of the CGLMP inequality (known as the Tsirelson's bound) can be derived from the complementarity of quantum correlations  --- the property, which was previously known only for correlations between many qubits \cite{BellComplementarity}.
Further we analyze the maximal violation of a three-qutrit inequality \cite{AcinChen04} from the perspective of its corresponding six-qubit Bell operator. We found a similar structure of maximally nonclassical states for this inequality, however the form of the inequality itself is much more complicated in this case.

\section{Single qutrit operators as symmetric two-qubit operators}

The linear space of qutrit operators, that is the space of matrices from $M_9(\mathbb C)$ has the same (complex) dimension of $9$ as the space of symmetric two-qubit operators, namely the operators from $\mathrm{Sym}(M_2(\mathbb C)\otimes M_2(\mathbb C))$. This shows, that the two spaces are isomorphic, and their elements are in one-to-one correspondence. This is a well known fact which is commonly used in the quantum theory of angular momentum. To get a deeper understanding of the isomorphism, let us proceed within a physically motivated approach. When discussing the operators of some quantum system, one is often interested in expressing them in an Hermitian operator basis. For finite dimensional quantum systems such a canonical Hermitian basis is the one consisting of Gell-Mann matrices \cite{Krammer08} --- Hermitian generators of the special unitary group $\mathrm{SU}(N)$, extended by the identity matrix. It turns out that in the case of qutrits, each Gell-Mann matrix can be expressed in more convenient way as a function of spin-$1$ matrices $\tilde{S_x},\tilde{S_y},\tilde{S_z}$, its squares $\tilde{S_x^2},\tilde{S_y^2},\tilde{S_z^2}$, and anticommutators $\{\tilde{S_x},\tilde{S_y}\}, \{\tilde{S_y},\tilde{S_z}\}, \{\tilde{S_x},\tilde{S_z}\}$ \cite{Krammer08},\cite{LM13} (from now on we will denote any qutrit operators and states with a tilde, to distinguish them from the qubit ones). The relation between Gell-Mann matrices and spin operators can be reversed, and one can get Hermitian basis consisting of spin-$1$ operators \cite{Kurzynski16} $\tilde{S_x},\tilde{S_y},\tilde{S_z}$,
shifted squares of the spin operators:
\begin{eqnarray}
\tilde{S_x^2} &=& \openone - (\tilde{S_x})^2,\nonumber\\
\tilde{S_y^2} &=& \openone - (\tilde{S_y})^2,\nonumber\\
\tilde{S_z^2} &=& \openone - (\tilde{S_z})^2,
\label{Spin32}
\end{eqnarray}
and all possible anticommutators:
\begin{eqnarray}
\tilde{A_x} &=& \tilde{S_z} \tilde{S_y} + \tilde{S_y} \tilde{S_z},\nonumber\\
\tilde{A_y} &=& \tilde{S_x} \tilde{S_z} + \tilde{S_z} \tilde{S_x},\nonumber\\
\tilde{A_z} &=& \tilde{S_x} \tilde{S_y} + \tilde{S_y} \tilde{S_z}.
\label{Spin33}
\end{eqnarray}
Note that the spin basis is not unique --- one can take as $\tilde{S_x},\tilde{S_y},\tilde{S_z}$ any set of Hermitian matrices fulfilling the spin commutation relations $[\tilde{S_x},\tilde{S_y}]=i \tilde{S_z}$ (and cyclic permutations of $x,y,z$). Two typical choices, to which we further refer are:
\begin{eqnarray}
&\tilde{S_x}=\frac{1}{\sqrt{2}}\left(\begin{array}{ccc}
0 & 1 & 0\\
1 & 0 & 1\\
0 & 1 & 0\\
\end{array}
\right),
\tilde{S_y}=\frac{1}{\sqrt{2}}\left(\begin{array}{ccc}
0 & -i & 0\\
i & 0 & -i\\
0 & i & 0\\
\end{array}
\right),&\nonumber\\
&\tilde{S_z}=\left(\begin{array}{ccc}
1 & 0 & 0\\
0 & 0 & 0\\
0 & 0 & -1\\
\end{array}
\right)&,
\label{SpinBasis1}
\end{eqnarray}
and:
\begin{eqnarray}
&\tilde{S_x}=\left(\begin{array}{ccc}
0 & 0 & 0\\
0 & 0 & -i\\
0 & i & 0\\
\end{array}
\right),
\tilde{S_y}=\left(\begin{array}{ccc}
0 & 0 & -i\\
0 & 0 & 0\\
i & 0 & 0\\
\end{array}
\right),&\nonumber\\
&\tilde{S_z}=\left(\begin{array}{ccc}
0 & i & 0\\
-i & 0 & 0\\
0 & 0 & 0\\
\end{array}
\right)&.
\label{SpinBasis2}
\end{eqnarray}
For the sake of convenience let us rename the spin basis elements as follows:
\be
\begin{array}{lll}
\gamma_1=\tilde{S_x}, & \gamma_2=\tilde{S_y}, & \gamma_3=\tilde{S_z},\\
\gamma_4=\openone - (\tilde{S_x})^2, & \gamma_5=\openone - (\tilde{S_y})^2, & \gamma_6=\openone - (\tilde{S_z})^2,\\
\gamma_7=\tilde{A_x}, & \gamma_8=\tilde{A_y}, & \gamma_9=\tilde{A_z}.\\
\end{array}
\label{SpinG}
\ee
Then any $n$-qutrit operator $\hat B\in M^3(\mathbb C)^{\otimes n}$ can be decomposed into a tensor form, by finding its coefficients in the product spin basis \eqref{SpinG} as follows:
\be
&&\hat B=\sum_{i_1,\ldots ,i_n}B_{i_1,\ldots , i_n}(\gamma_{i_1}\otimes\ldots\otimes\gamma_{i_n}) \nonumber\\
&&B_{i_1,\ldots , i_n}=\frac{\Tr(\hat B (\gamma_{i_1}\otimes\ldots\otimes\gamma_{i_n}))}{\Tr\left((\gamma_{i_1}\otimes\ldots\otimes\gamma_{i_n})^2\right)},
\label{BtoSpin1}
\ee
where the denominator compensates the fact that the trace norms of the basis elements may be different.
However, the spin-$1$ operators can be represented in another way, which comes from the composition of two spin-half systems. From the point of view of the (non-relativistic) symmetries of a physical system, the spin of a particle determines how its state vector reacts on rotations of the physical space $\mathbb R^3$, or saying more mathematically --- under which representation of the rotation group the state transforms. If we compose two particles of spin-half, therefore each transforming under rotation as a two-dimensional spinor, the state space of the composite system contains two invariant subspaces with respect to three dimensional rotations: the fully symmetric space (of complex dimension $3$), the vectors of which transform under rotations as if they correspond to spin-$1$ particle, and the invariant one dimensional subspace spanned by the \emph{singlet state}. From the composition rule for generators of the tensor product of arbitrary transformations, it follows that the effective spin-$1$ operators acting on the symmetric subspace of two-qubits have the following direct-sum form:
\begin{eqnarray}
\tilde{S_x} \mapsto S_x &=& \frac{1}{2} (\openone \otimes X + X \otimes \openone )\equiv \delta_1, \nonumber \\
\tilde{S_y} \mapsto S_y &=& \frac{1}{2} (\openone \otimes Y + Y \otimes \openone )\equiv \delta_2, \nonumber \\
\tilde{S_z} \mapsto S_z &=& \frac{1}{2} (\openone \otimes Z + Z \otimes \openone )\equiv \delta_3, 
\label{Spin43}
\end{eqnarray}
where $X,Y,Z$ are standard qubit Pauli matrices (or their unitarily rotated equivalents), and $\delta_i$ is a shorthand notation analogous to \eqref{SpinG}. The other spin-$1$ basis operators \eqref{Spin32}--\eqref{Spin33} are transformed as follows: 
\begin{eqnarray}
\tilde{S_x^2} \mapsto S_x^2 &=&\delta_4= \frac{1}{4} (\openone - X \otimes X + Y \otimes Y + Z \otimes Z) \nonumber\\
&=& \left| \Phi^- \right\rangle \left\langle  \Phi^-\right|, \nonumber \\
\tilde{S_y^2} \mapsto S_y^2 &=&\delta_5= \frac{1}{4} (\openone + X \otimes X - Y \otimes Y + Z \otimes Z) \nonumber\\
&=& \left| \Phi^+ \right\rangle \left\langle  \Phi^+\right|, \nonumber \\
\tilde{S_z^2} \mapsto S_z^2 &=&\delta_6= \frac{1}{4} (\openone + X \otimes X + Y \otimes Y - Z \otimes Z) \nonumber\\
&=& \left| \Psi^+ \right\rangle \left\langle  \Psi^+\right|, \nonumber \\
\tilde{A_x} \mapsto A_x &=&\delta_7= \frac{1}{2} (Y \otimes Z + Z \otimes Y), \nonumber \\
\tilde{A_y} \mapsto A_y &=&\delta_8= \frac{1}{2} (X \otimes Z + Z \otimes X), \nonumber \\
\tilde{A_z} \mapsto A_z &=&\delta_9= \frac{1}{2} (X \otimes Y + Y \otimes X),
\label{Spin44}
\end{eqnarray}
where we used the standard notation $\ket{\Phi^-},\ket{\Phi^+},\ket{\Psi^+}$ for the symmetric Bell states of two qubits.
Here, the $S_k^2$ operators are obtained by using the formula:
\begin{equation}
S_k^2 = \openone_{\textrm{sym}} - (S_k)^2,
\end{equation}
where $\openone_{\textrm{sym}} = \openone - \left| \Psi^- \right\rangle \left\langle  \Psi^-\right|$ is the identity matrix on a symmetric subspace of two qubits and $S_k^2 = \frac{1}{2} (\openone \otimes \openone + K \otimes K)$ ($K$ is k-th Pauli matrix).
Using the above defined representation of a spin-$1$ basis, we can decompose any $n$-qutrit operator as a symmetric $2n$-qubit operator:
\be
\hat B=\sum_{i_1,\ldots ,i_n}B_{i_1,\ldots , i_n}(\delta_{i_1}\otimes\ldots\otimes\delta_{i_n}),
\label{BtoSpin2}
\ee
where the expansion coefficients are the same as in \eqref{BtoSpin1}.
\color[rgb]{0,0,0}
So far we discussed the relations between the spin-$1$ operators, however the transformation rule for states is equally important. Let us denote some fixed qutrit orthonormal basis as $\{\tilde{\ket{0}},\tilde{\ket{1}},\tilde{\ket{2}}\}$. If we map the qutrit operators to symmetric two-qubit operators, the corresponding qutrit basis states are mapped to the basis $\{e_1,e_2,e_3\}$ consisting of some symmetric states of two qubits. These states have to transform in the same way under the action of corresponding spin operators as the $\{\tilde{\ket{0}},\tilde{\ket{1}},\tilde{\ket{2}}\}$. If we fix the operator representations for $\{\tilde{S_x},\tilde{S_y},\tilde{S_z}\}=\{\gamma_1,\gamma_2,\gamma_3\}$, the basis for Pauli matrices in \eqref{Spin43}, and the qutrit standard basis $\{\tilde{\ket{0}},\tilde{\ket{1}},\tilde{\ket{2}}\}$, then the new qutrit basis $e_i$ can be derived from the equality of the matrix elements:
\be
\forall_{i,k=0,1,2}\,\,\forall_{j=1,\ldots,9}\,\,\, \tilde{\bra{i}}\gamma_j\tilde{\ket{k}}=\bra{e_i}\delta_j\ket{e_k}.
\label{SpinState}
\ee
Let us now fix the standard representation for Pauli matrices in \eqref{Spin43} and the standard qutrit basis for $\{\tilde{\ket{0}},\tilde{\ket{1}},\tilde{\ket{2}}\}$. Then the choice of spin-$1$ operators in the form \eqref{SpinBasis1} implies the following transformation rules for states:
\be
&&\tilde{\ket{0}}\mapsto \ket{00},\nonumber\\
&&\tilde{\ket{1}}\mapsto\frac{1}{\sqrt{2}}(\ket{01}+\ket{10}),\nonumber\\
&&\tilde{\ket{2}}\mapsto \ket{11},
\label{NEW3to2}
\ee
whereas the choice of the set \eqref{SpinBasis2} implies the following:
\be
&&\tilde{\ket{0}}\mapsto\frac{i}{\sqrt{2}}(\ket{00}-\ket{11}),\nonumber\\
&&\tilde{\ket{1}}\mapsto\frac{1}{\sqrt{2}}(\ket{00}+\ket{11}),\nonumber\\
&&\tilde{\ket{2}}\mapsto\frac{i}{\sqrt{2}}(\ket{01}+\ket{10}).
\label{OLD3to2}
\ee

\section{CGLMP Bell operator in the symmetric two-qubit operators representation}

The CGLMP inequality in its simplest version \cite{CGLMPorigin, CGLMPacin} is a Bell inequality for two observers $\mathcal A$ and $\mathcal B$, each having two measurement settings $\{A_1,A_2\}$, $\{B_1,B_2\}$ with $3$ outcomes, labeled as $\{0,1,2\}$. The inequality is originally presented as a constraint on a linear function for two-outcome probabilities \cite{CGLMPacin}:
\be
&&I_3=P(A_1=B_1)+P(A_2+1=B_1)+P(A_2=B_2)+\nonumber\\
&&P(A_1=B_2)-P(A_1=B_1-1)-P(A_2=B_1)\nonumber\\
&&-P(A_2=B_2-1)-P(A_1-1=B_2).
\label{originCGLMP}
\ee
In the case of classical probabilities (admitting a joint probability distribution) the above value is bounded from both sides as follows \cite{Chen06}:
\be
-4&\leq& I_3\leq 2.
\label{CGLMPbound}
\ee
The effective way to discuss a maximal quantum violation of a Bell inequality by some quantum system is the method of a Bell operator \cite{Braunstein92}. This method relies on evaluation of the value of the body of a Bell inequality (in our case $I_3$ \eqref{originCGLMP}) for a given quantum state $\rho$ and given measurement settings $\{\hat A_1,\hat A_2,\hat B_1,\hat B_2\}$  by the Born rule:
\be
I_3(\rho)=\Tr(\hat B(\hat A_1,\hat A_2,\hat B_1,\hat B_2) \rho).
\label{BellOp}
\ee
The Bell operator $\hat B$ can be found by summing the single-run probabilities:
\be
P(A_k=i,B_m=j)&=&\Tr((\proj{i}\otimes\proj{j})\rho),
\label{BellOp1}
\ee
using the relation $P(A=B+k)=\sum_{j=0}^{2}P(A=j,B=j+k\textrm{ mod } 3)$ \cite{CGLMPorigin}. 
The maximal quantum violation of a Bell inequality for a given set of settings equals to the largest eigenvalue of the Bell operator, and the optimal state is its corresponding eigenstate.
The CGLMP \eqref{originCGLMP} Bell operator for the optimal settings has the following form \cite{CGLMPacin}:
\begin{equation}
\hat B = \left(\begin{array}{ccccccccc}
    0& 0& 0& 0& \frac{2}{\sqrt{3}}& 0& 0& 0& 2 \\
    0& 0& 0& 0& 0& \frac{2}{\sqrt{3}}& 0& 0& 0 \\
    0& 0& 0& 0& 0& 0& 0& 0& 0 \\
	  0& 0& 0& 0& 0& 0& 0& \frac{2}{\sqrt{3}}& 0 \\
	  \frac{2}{\sqrt{3}}& 0& 0& 0& 0& 0& 0& 0& \frac{2}{\sqrt{3}}\\
	  0& \frac{2}{\sqrt{3}}& 0& 0& 0& 0& 0& 0& 0 \\
	  0& 0& 0& 0& 0& 0& 0& 0& 0 \\
	  0& 0& 0& \frac{2}{\sqrt{3}}& 0& 0& 0& 0& 0 \\
	  2& 0& 0& 0& \frac{2}{\sqrt{3}}& 0& 0& 0& 0\\
		\end{array}
		\right).
		\label{BellOpC}
\end{equation}
The highest eigenvalue of the Bell operator equals $1+\sqrt{\frac{11}{3}}\approx 2.915$ and the corresponding eigenstate is:
\be
\ket{\psi_{max}}=a \tilde{|00 \rangle} + b \tilde{|11 \rangle} + a \tilde{|22\rangle},
\label{CGLMPmaxState}
\ee
where $a = \frac{5 \sqrt{3} + 3 \sqrt{11}}{\sqrt{462 + 78 \sqrt{33}}} \approx 0.617$ and $b = \frac{ 9 + \sqrt{33}}{\sqrt{462 + 78 \sqrt{33}}} \approx 0.489$.
In the case of a maximally entangled two-qutrit state $\ket{\psi_{ME}}=\frac{1}{\sqrt{3}}(\tilde{\ket{00}}+\tilde{\ket{11}}+\tilde{\ket{22}})$ the inequality \eqref{CGLMPbound} is violated slightly less: $I_3(\ket{\psi_{ME}})\approx 2.873$, leading to an inconsistency between maximal quantum entanglement and maximal nonclassicality in terms of violation of the optimal Bell inequality. 

\begin{figure}
\includegraphics[width=0.80\columnwidth]{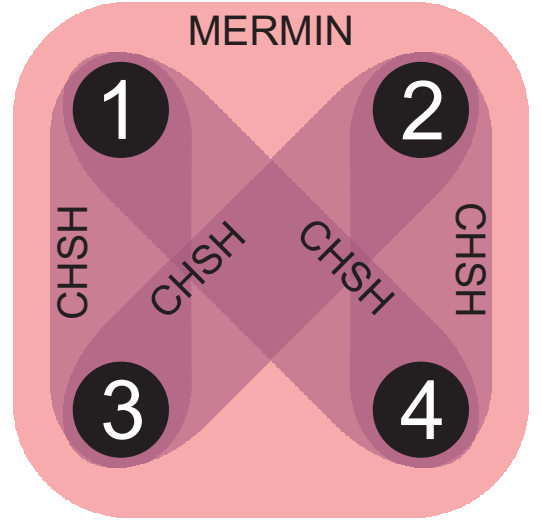}
\caption{Schematic presentation of a CGLMP Bell operator \eqref{BellOp2q} in the $4$-qubit representation. The operator consists of five parts, the one corresponding to a $4$-qubit Mermin's inequality, and $4$ corresponding to CHSH inequalities for all pairs of qubits.}
\label{BellOperators}
\end{figure}

In order to resolve the paradox, we transform the Bell operator \eqref{BellOpC} to spin-$1$ bases. Using the transformation \eqref{BtoSpin1} we obtain the following tensor form of the CGLMP Bell operator:
\begin{eqnarray}
&&\hat B=\frac{2}{\sqrt{3}} (\tilde{S_x} \otimes \tilde{S_x} - \tilde{S_y} \otimes \tilde{S_y})+\tilde{S_x^2} \otimes \tilde{S_x^2}\nonumber\\
&&+ \tilde{S_y^2} \otimes \tilde{S_y^2} - \tilde{S_x^2} \otimes \tilde{S_y^2}-  \tilde{S_y^2} \otimes \tilde{S_x^2}-\tilde{A_z} \otimes \tilde{A_z}.
\label{BellOpSpin1}
\end{eqnarray}
%\begin{eqnarray}
%&&\hat B_{new}=\frac{1}{\sqrt{3}} (A_x \otimes A_x + A_z \otimes A_z + S_x \otimes S_z + S_z \otimes S_x \nonumber\\
%&&- A_x \otimes A_z - A_z \otimes A_x - S_x \otimes S_x - S_z \otimes S_z)\nonumber\\
%&&-S_y \otimes S_y + A_y \otimes A_y
%\label{newBellOpSpin1}
%\end{eqnarray}
Note that since $\tilde{A_z}$ is defined as the anticommutator of $\tilde{S_x}$ and $\tilde{S_y}$, the above operator is build solely with the spin operators corresponding to $x-y$ plane. Further we transform the Bell operator to the symmetric qubit basis using \eqref{BtoSpin2}:
\begin{eqnarray}
\hat B=\frac{1}{4} \bigg(\frac{2}{\sqrt{3}} \Big\{
   X\otimes \openone\otimes X\otimes \openone 
 &-& Y\otimes \openone\otimes Y\otimes \openone  \nonumber\\
 + X\otimes \openone\otimes \openone\otimes X 
 &-& Y\otimes \openone\otimes \openone\otimes Y  \nonumber\\
 + \openone\otimes X\otimes X\otimes \openone 
 &-& \openone\otimes Y\otimes Y\otimes \openone  \nonumber\\
 + \openone\otimes X\otimes \openone\otimes X 
 &-& \openone\otimes Y\otimes \openone\otimes Y\Big\} \nonumber\\
						+ X\otimes X\otimes X\otimes X &+& Y\otimes Y\otimes Y\otimes Y \nonumber\\- Y\otimes Y\otimes X\otimes X &-&
    X\otimes X\otimes Y\otimes Y \nonumber\\
		- Y\otimes X\otimes Y\otimes X &-& Y\otimes X\otimes X\otimes Y \nonumber\\ - X\otimes Y\otimes X\otimes Y &-& 
   X\otimes Y\otimes Y\otimes X\bigg).\nonumber\\
	\label{BellOp2q}
\end{eqnarray}
The above operator as a $4$-qubit operator can be now related to known Bell operators for qubit Bell inequalities. Indeed, the first part of the operator (denoted in $\{\}$ braces) corresponds to Bell operators for CHSH inequalities \cite{CHSH, Braunstein92} for all four pairs of qubits, whereas the second part corresponds to a Bell operator of a four-qubit Mermin inequality \cite{Mermin90}. The structure of the Bell operator \eqref{BellOp2q} is schematically presented in the Fig. \ref{BellOperators}.
Since the Mermin inequality is maximally violated by a GHZ state, whereas CHSH inequalities are maximally violated by two-qubit Bell states, one can expect that the eigenvector corresponding to the largest eigenvalue of \eqref{BellOp2q} is a superposition of a GHZ state and two Bell states (due to the monogamy of violation of CHSH inequalities \cite{Toner06}, only two of the four CHSH inequalities, corresponding to separate pairs of qubits, can be violated maximally at the same time):
\begin{equation}
|\psi (p) \rangle = \sqrt{p} |GHZ\rangle + \sqrt{1-p} |\psi^+\rangle|\psi^+\rangle,
\label{supGHZBell}
\end{equation}
for which: 
\be
{\rm Tr}(\hat B |\psi(p)\rangle\langle\psi(p)|) =  2p + 4 \sqrt{\frac{2 p (1-p)}{3}}.
\label{EigB2q}
\ee
It turns out that the quantity \eqref{EigB2q} is maximized for $p_{\textrm{max}}=\frac{1}{22}(11+\sqrt{33})\approx 0.761$, and $\ket{\psi(p_{\textrm{max}})}$ is a representation of a state \eqref{CGLMPmaxState}, which maximally violates CGLMP inequality, in symmetric-two-qubit basis. Indeed, by taking the set of transformations \eqref{NEW3to2} one easily obtains the state \eqref{CGLMPmaxState} from $\ket{\psi(p_{\textrm{max}})}$. The maximally entangled state of two qutrits corresponds via transformations \eqref{NEW3to2} to the state $\ket{\psi(\frac{2}{3})}$:
\begin{eqnarray}
\label{psi4}
&&|\psi\left(\tfrac{2}{3}\right)\rangle=\sqrt{\frac{2}{3}}|GHZ\rangle + \sqrt{\frac{1}{3}} |\psi^+\rangle |\psi^+\rangle\nonumber\\
&&\equiv \frac{1}{\sqrt{3}}(\tilde{|00\rangle}+\tilde{|11\rangle}+\tilde{|22\rangle}).
\end{eqnarray}
The symmetric two-qubit form of a CGLMP Bell operator \eqref{BellOp2q} explains, why the maximally entangled state of two qutrits does not give rise to a maximal violation: the CGLMP inequality is violated by the state from  the family \eqref{supGHZBell}, and the optimal $p$, which maximizes the violation \eqref{EigB2q}, is determined by the structure constants of the operator \eqref{BellOp2q}. In this representation the maximally entangled state of two-qutrits seems to be suboptimal. 

One additional comment is necessary here. The Bell operator \eqref{BellOp2q}, which represents a CGLMP inequality, does not give rise to a four-qubit Bell inequality. Namely, the corresponding correlation-based Bell inequality for $4$ parties, $2$ settings and $2$ outcomes:
\begin{eqnarray}
&&1-\frac{4}{\sqrt{3}}\leq \frac{1}{4}\bigg(\frac{2}{\sqrt{3}} \Big(A_1 C_1 - A_2 C_2  + A_1 D_1 - A_2 D_2 \nonumber\\
&&+ B_1 C_1 - B_2 C_2 + B_1 D_1 - B_2 D_2\Big) \nonumber \\
&&+ A_1 B_1 C_1 D_1 + A_2 B_2 C_2 D_2 - A_1 B_1 C_2 D_2 - A_1 B_2 C_1 D_2 \nonumber \\
&&- A_2 B_1 C_1 D_2 - A_1 B_2 C_2 D_1 - A_2 B_1 C_2 D_1 - 
    A_2 B_2 C_1 D_1\bigg) \nonumber \\ &&\leq 1+ \frac{4}{\sqrt{3}},
		\label{BellIn4q}
		\end{eqnarray}
is not violated by any quantum state. Therefore, all the above considerations of the Bell operator \eqref{BellOp2q} within the symmetric-two-qubit representation must be treated as a tool for analyzing the physical properties of qutrits.

Finally we show, that the symmetric-two-qubit representation of a CGLMP Bell operator allows for deriving the maximal quantum violation (the Tsirelson bound) from complementarity of quantum correlations.
Let us first introduce the following notation:
\begin{eqnarray}
\alpha &=& \langle X\otimes \openone\otimes X\otimes \openone \rangle
=\langle X\otimes \openone\otimes \openone\otimes X\rangle \\ \nonumber
&=&\langle \openone\otimes X\otimes X\otimes \openone \rangle
=\langle \openone\otimes X\otimes \openone\otimes X \rangle \\
\beta &=& \langle Y\otimes \openone\otimes \openone\otimes Y \rangle 
=\langle Y\otimes \openone\otimes Y\otimes \openone \rangle \\ \nonumber
&=&\langle \openone\otimes Y\otimes Y\otimes \openone \rangle 
=\langle \openone\otimes Y\otimes \openone\otimes Y \rangle \\
\tau &=& \langle Y\otimes X\otimes Y\otimes X \rangle
=\langle Y\otimes X\otimes X\otimes Y\rangle\\ \nonumber
&=&\langle X\otimes Y\otimes X\otimes Y\rangle
=\langle X\otimes Y\otimes Y\otimes X \rangle \\
\epsilon &=& \langle X\otimes X\otimes Y\otimes Y\rangle=\langle Y\otimes Y\otimes X\otimes X\rangle  \\
1 &=& \langle X\otimes X\otimes X\otimes X \rangle= \langle Y\otimes Y\otimes Y\otimes Y \rangle
\end{eqnarray}
Then the mean value of the Bell operator \eqref{BellOp2q} reads:
\begin{equation}
\langle B \rangle  = \frac{1}{4} \left(\frac{8}{\sqrt{3}} (\alpha - \beta) - 4 \tau - 
     2 \epsilon + 2\right)
\end{equation}	
We follow the approach of \cite{BellComplementarity}, in which one finds sets of mutually maximally anticommuting operators. It is shown, that squares of mean values of such operators are upper-bounded by $1$ due to the complementarity of correlations. It can be easily shown that the operator sets $\{\alpha,\beta,\epsilon\}$, $\{\alpha,\tau\}$ and $\{\beta,\tau\}$ are maximally anticommuting, therefore the following constraints are valid:
		\begin{eqnarray}
		&&\alpha^2 + \beta^2 + \epsilon^2 \leq 1 ,\nonumber\\
	&&\alpha^2 + \tau^2 \leq 1,\nonumber\\
   &&\beta^2 + \tau^2 \leq 1.
	\label{cons1}
 					\end{eqnarray}
We add the following constraint:
\be
\epsilon - 2 \tau \leq 1,
\label{cons2}
\ee
which follows from the nonnegativity of the expectation value of any projector. Note that the following expression:
\begin{eqnarray}
\Pi&=&2(\openone^{\otimes 4}) - X \otimes X \otimes X \otimes X  - X \otimes X \otimes Y \otimes Y \nonumber \\ &+&  Y \otimes X \otimes Y \otimes X +  Y \otimes X \otimes X \otimes Y,
\end{eqnarray}
is an operator with eigenvalues $4$ and $0$, therefore it is proportional to a projector. The expectation value of this expression:
$\langle\Pi\rangle=1-\epsilon+2\tau$ is greater than zero and hence we get (\ref{cons2}).

We maximize the mean of the Bell operator under the constraints \eqref{cons1} and \eqref{cons2}:
\begin{equation}
\max_{\alpha, \beta, \tau,  \epsilon } \langle B \rangle = \frac{1}{4} \left(4 + 4 \sqrt{\frac{11}{3}}\right) \approx 2.915,
\end{equation}
which exactly gives the maximal quantum violation.

\section{Maximal entanglement \emph{vs} maximal  non-classicality beyond CGLMP}

The two-qutrit CGLMP inequality \cite{CGLMPorigin} found its direct generalization to the case of three higher dimensional parties \cite{AcinChen04, Chen08}. Especially interesting is the case of a three-qutrit inequality \cite{AcinChen04}:
\be
&&P(A_1+B_1+C_1=0)+P(A_1+B_2+C_2=1)+\nonumber\\
&&P(A_2+B_1+C_2=0)+P(A_2+B_2+C_1=1)+\nonumber\\
&&2P(A_2+B_2+C_2=0)-P(A_2+B_1+C_1=2)-\nonumber\\
&&P(A_1+B_2+C_1=2)-P(A_1+B_1+C_2=2)\leq 3,\nonumber\\
\label{3CGLMP}
\ee
the features of which are very similar to the original CGLMP: it is tight, and its maximal violation of $4.372$ arises for a slightly non-maximally entangled state \eqref{CGLMPmaxState}
\be
\ket{\psi_{max}}=a \tilde{|000 \rangle} + b \tilde{|111 \rangle} + a \tilde{|222\rangle},
\label{MaxState3q}
\ee
where $a = \frac{5 \sqrt{3} + 3 \sqrt{11}}{\sqrt{462 + 78 \sqrt{33}}} \approx 0.617$ and $b = \frac{ 9 + \sqrt{33}}{\sqrt{462 + 78 \sqrt{33}}} \approx 0.489$.

For a maximally entangled three-qutrit state one obtains the violation of $4.333$. Using analogous techniques like in CGLMP case we can derive the Bell operator for this inequality and translate it to two-qubit symmetric basis \eqref{BtoSpin2}. Although the form of the Bell operator in the spin-$1$ basis is very complicated in this case (see Appendix \ref{App1}), we can easily discuss the form of an optimal state giving the maximal violation in the $6$-qubit representation. In full analogy to \eqref{supGHZBell}, the maximally entangled state of three qutrits translates under transformations \eqref{NEW3to2} to:
\be
&&\frac{1}{\sqrt{3}}(\ket{\tilde{0}}^{\otimes 3}+\ket{\tilde{0}}^{\otimes 3}+\ket{\tilde{0}}^{\otimes 3})\mapsto \sqrt{\frac{2}{3}} |GHZ\rangle + \sqrt{\frac{1}{3}} |\psi^+\rangle^{\otimes 3}.\nonumber\\
%&&0.8165|GHZ\rangle + 0.5774 |\psi^+\rangle^{\otimes 3}.
\ee
If we optimize the violation over all possible superpositions of $6$-qubit GHZ state and $3$ Bell states:
\be
\ket{\psi(p)}=\sqrt{p} |GHZ\rangle + \sqrt{1-p} |\psi^+\rangle^{\otimes 3},
\ee
we obtain the maximal value of $4.345$ for $p\approx 0.845$, which is slightly larger, but still suboptimal. It turns out, that in order to find a maximal violation we have to search over the following family of states:
\be
\ket{\psi(p,\theta)}=\sqrt{p} \left(\sin(\theta)|0\rangle^{\otimes 6}+\cos(\theta)|1\rangle^{\otimes 6}\right) + \sqrt{1-p} |\psi^+\rangle^{\otimes 3},\nonumber\\
\ee 
which is a superposition of a generalized GHZ state (with unequal weights) and the product of $3$ Bell states. The maximal violation is attained for $\theta\approx 0.870$ and  $p\approx 0.841$, which reproduces the state \eqref{MaxState3q}. 

In this case the interpretation of the optimal form of a $6$-qubit equivalent of a $3$-qutrit state is not so straightforward as in the case of CGLMP. 

\section{Conclusions}
In this work we discussed various aspects of a nonclassicality of qutrit states in terms of violation of tight Bell inequalities. We introduced a new method of analyzing the maximal violation of a Bell inequality by transforming its Bell operator to a local basis of symmetric two-qubit operators. In this way the analysis of a Bell inequality for $n$ qutrits is translated into the analysis of a corresponding Bell inequality for $2n$ qubits. Using this method in the case of a CGLMP inequality we resolved the paradox of a maximal violation by a non-maximally entangled two-qutrit state. Moreover, we were able to derive the Tsirelson bound for the CGLMP inequality solely from the complementarity of correlations, which has never been observed before for correlations between qutrits. 

\section{Acknowledgements}
MM, PK, WL and AK are supported by NCN Grant No. 2014/14/M/ST2/00818. KK is supported by NCN Grant No. 2012/05/E/ST2/02352.
\appendix
\section{The $3$-qutrit inequality \cite{AcinChen04} Bell operator in spin-$1$ bases}
\label{App1}

Here, we list the sets of elements of the Bell operator for the $3$-qutrit Bell inequality \eqref{3CGLMP} in the spin-1 basis \eqref{SpinG} using the rule \eqref{BtoSpin1}:
\be
&&B_{[122]} = B_{[177]} = B_{[279]} = - \frac{3}{8 \sqrt{2}},\nonumber \\
&&B_{[127]} = B_{[228]} = B_{[778]} = - \frac{1}{8 \sqrt{2}},\nonumber \\
&&B_{444}   = B_{[455]} = -\frac{1}{4}, \nonumber \\
&&B_{[599]} = -B_{[499]} = - \frac{1}{2}, \nonumber \\
&&B_{[334]} = B_{[335]} = \frac{1}{4}, \nonumber \\
&&B_{888}   = B_{[118]} = \frac{1}{8 \sqrt{2}},\nonumber \\
&&B_{111}   = B_{[188]} = \frac{3}{8 \sqrt{2}},\nonumber \\
&&B_{555}   = B_{[445]} = \frac{3}{4}, \nonumber \\
&&B_{444}   = 1,
\ee
where the indices denote the subscript indices of the corresponding operators $\gamma_i$ \eqref{SpinG} and $[...]$ denotes all possible permutations of given indices.

After transformation into the $6$-qubit representation (according to the formula {\eqref{BtoSpin2}} )  the Bell operator of the inequality \eqref{3CGLMP} reads:
\be
&& B_{[(0 1)(0 2)(0 2)]} = B_{[(0 1)(23)(23)]} = B_{[(13)(0 2)(23)]} = -\frac{3}{\sqrt{2}},\nonumber \\
&& B_{[(0 1)(0 2)(23)]} = B_{[(0 1)(0 1)(13)]} = B_{[(0 2)(0 2)(13)]} = -\frac{1}{\sqrt{2}},\nonumber \\
&& B_{[(11)(12)(12)]} = -4,\nonumber \\
&& B_{2 2 2 2 2 2} = B_{[(11)(11)(22)]} = -3,\nonumber \\
&& B_{[(0 0)(11)(33)]} = B_{[(0 0)(22)(33)]} = -1,\nonumber \\
&& B_{[(0 0)(0 0)(11)]} = B_{[(0 0)(0 0)(22)]} = B_{[(0 0)(0 0)(33)]} \nonumber \\
&& = B_{[(0 0)(11)(11)]} = B_{[(0 0)(22)(22)]} = B_{[(0 0)(33)(33)]} \nonumber \\
&& = B_{[(0 0)(11)(22)]} = B_{[(11)(33)(33)]} = B_{[(22)(33)(33)]} \nonumber \\
&& = B_{3 3 3 3 3 3} = 1, \nonumber \\
&& B_{[(0 0)(0 3)(0 3)]} = B_{[(0 3)(0 3)(33)]} = 2,\nonumber \\
&& B_{0 0 0 0 0 0} = B_{[(0 0)(33)(33)]} = 3,\nonumber \\
&& B_{[(12)(12)(22)]} = 4,\nonumber \\
&& B_{1 1 1 1 1 1} = B_{[(11)(22)(22)]} = 5,\nonumber\\
&& B_{[(0 1)(0 1)(0 1)]} = B_{[(0 1)(13)(13)]} = \frac{3}{\sqrt{2}},\nonumber \\
&& B_{[(0 1)(0 1)(23)]} = B_{[(13)(13)(13)]} = \frac{1}{\sqrt{2}}.
\label{6qubitBell}
\ee
We use a modified notation in which the subscript numbers denote three Pauli matrices (the convention about the indices is: $1=X$, $2=Y$, $3=Z$) and the identity matrix (index $0$).
Here, all elements should be normalized by the factor $\frac{1}{16}$. $[...]$ denotes the set of all possible permutations of a given subset and a pair $(L,K)$ describes combination $L \otimes K$ or $K \otimes L$ (of Pauli matrices or identity) traveling always together through $2$-partite subsystems of indices (it means, that the combination $(L,K)$ is always measured on the first and second, third and fourth or fifth and sixth subsystem respectively).

%merlin.mbs apsrev4-1.bst 2010-07-25 4.21a (PWD, AO, DPC) hacked
%Control: key (0)
%Control: author (72) initials jnrlst
%Control: editor formatted (1) identically to author
%Control: production of article title (-1) disabled
%Control: page (0) single
%Control: year (1) truncated
%Control: production of eprint (0) enabled
%

%\bibliographystyle{apsrev4-1}
%\bibliography{CGLMP}

\begin{thebibliography}{23}%
\makeatletter
\providecommand \@ifxundefined [1]{%
 \@ifx{#1\undefined}
}%
\providecommand \@ifnum [1]{%
 \ifnum #1\expandafter \@firstoftwo
 \else \expandafter \@secondoftwo
 \fi
}%
\providecommand \@ifx [1]{%
 \ifx #1\expandafter \@firstoftwo
 \else \expandafter \@secondoftwo
 \fi
}%
\providecommand \natexlab [1]{#1}%
\providecommand \enquote  [1]{``#1''}%
\providecommand \bibnamefont  [1]{#1}%
\providecommand \bibfnamefont [1]{#1}%
\providecommand \citenamefont [1]{#1}%
\providecommand \href@noop [0]{\@secondoftwo}%
\providecommand \href [0]{\begingroup \@sanitize@url \@href}%
\providecommand \@href[1]{\@@startlink{#1}\@@href}%
\providecommand \@@href[1]{\endgroup#1\@@endlink}%
\providecommand \@sanitize@url [0]{\catcode `\\12\catcode `\$12\catcode
  `\&12\catcode `\#12\catcode `\^12\catcode `\_12\catcode `\%12\relax}%
\providecommand \@@startlink[1]{}%
\providecommand \@@endlink[0]{}%
\providecommand \url  [0]{\begingroup\@sanitize@url \@url }%
\providecommand \@url [1]{\endgroup\@href {#1}{\urlprefix }}%
\providecommand \urlprefix  [0]{URL }%
\providecommand \Eprint [0]{\href }%
\providecommand \doibase [0]{http://dx.doi.org/}%
\providecommand \selectlanguage [0]{\@gobble}%
\providecommand \bibinfo  [0]{\@secondoftwo}%
\providecommand \bibfield  [0]{\@secondoftwo}%
\providecommand \translation [1]{[#1]}%
\providecommand \BibitemOpen [0]{}%
\providecommand \bibitemStop [0]{}%
\providecommand \bibitemNoStop [0]{.\EOS\space}%
\providecommand \EOS [0]{\spacefactor3000\relax}%
\providecommand \BibitemShut  [1]{\csname bibitem#1\endcsname}%
\let\auto@bib@innerbib\@empty
%</preamble>
\bibitem [{\citenamefont {Werner}(1989)}]{Werner89}%
  \BibitemOpen
  \bibfield  {author} {\bibinfo {author} {\bibfnamefont {R.~F.}\ \bibnamefont
  {Werner}},\ }\href {\doibase 10.1103/PhysRevA.40.4277} {\bibfield  {journal}
  {\bibinfo  {journal} {Phys. Rev. A}\ }\textbf {\bibinfo {volume} {40}},\
  \bibinfo {pages} {4277} (\bibinfo {year} {1989})}\BibitemShut {NoStop}%
\bibitem [{\citenamefont {Collins}\ \emph {et~al.}(2002)\citenamefont
  {Collins}, \citenamefont {Gisin}, \citenamefont {Linden}, \citenamefont
  {Massar},\ and\ \citenamefont {Popescu}}]{CGLMPorigin}%
  \BibitemOpen
  \bibfield  {author} {\bibinfo {author} {\bibfnamefont {D.}~\bibnamefont
  {Collins}}, \bibinfo {author} {\bibfnamefont {N.}~\bibnamefont {Gisin}},
  \bibinfo {author} {\bibfnamefont {N.}~\bibnamefont {Linden}}, \bibinfo
  {author} {\bibfnamefont {S.}~\bibnamefont {Massar}}, \ and\ \bibinfo {author}
  {\bibfnamefont {S.}~\bibnamefont {Popescu}},\ }\href {\doibase
  10.1103/PhysRevLett.88.040404} {\bibfield  {journal} {\bibinfo  {journal}
  {Phys. Rev. Lett.}\ }\textbf {\bibinfo {volume} {88}},\ \bibinfo {pages}
  {040404} (\bibinfo {year} {2002})}\BibitemShut {NoStop}%
\bibitem [{\citenamefont {Ac\'{\i}n}\ \emph {et~al.}(2004)\citenamefont
  {Ac\'{\i}n}, \citenamefont {Chen}, \citenamefont {Gisin}, \citenamefont
  {Kaszlikowski}, \citenamefont {Kwek}, \citenamefont {Oh},\ and\ \citenamefont
  {\ifmmode~\dot{Z}\else \.{Z}\fi{}ukowski}}]{AcinChen04}%
  \BibitemOpen
  \bibfield  {author} {\bibinfo {author} {\bibfnamefont {A.}~\bibnamefont
  {Ac\'{\i}n}}, \bibinfo {author} {\bibfnamefont {J.~L.}\ \bibnamefont {Chen}},
  \bibinfo {author} {\bibfnamefont {N.}~\bibnamefont {Gisin}}, \bibinfo
  {author} {\bibfnamefont {D.}~\bibnamefont {Kaszlikowski}}, \bibinfo {author}
  {\bibfnamefont {L.~C.}\ \bibnamefont {Kwek}}, \bibinfo {author}
  {\bibfnamefont {C.~H.}\ \bibnamefont {Oh}}, \ and\ \bibinfo {author}
  {\bibfnamefont {M.}~\bibnamefont {\ifmmode~\dot{Z}\else \.{Z}\fi{}ukowski}},\
  }\href {\doibase 10.1103/PhysRevLett.92.250404} {\bibfield  {journal}
  {\bibinfo  {journal} {Phys. Rev. Lett.}\ }\textbf {\bibinfo {volume} {92}},\
  \bibinfo {pages} {250404} (\bibinfo {year} {2004})}\BibitemShut {NoStop}%
\bibitem [{\citenamefont {Laskowski}\ \emph {et~al.}(2014)\citenamefont
  {Laskowski}, \citenamefont {Ryu},\ and\ \citenamefont {\.Zukowski}}]{LRZ14}%
  \BibitemOpen
  \bibfield  {author} {\bibinfo {author} {\bibfnamefont {W.}~\bibnamefont
  {Laskowski}}, \bibinfo {author} {\bibfnamefont {J.}~\bibnamefont {Ryu}}, \
  and\ \bibinfo {author} {\bibfnamefont {M.}~\bibnamefont {\.Zukowski}},\
  }\href {http://stacks.iop.org/1751-8121/47/i=42/a=424019} {\bibfield
  {journal} {\bibinfo  {journal} {Journal of Physics A: Mathematical and
  Theoretical}\ }\textbf {\bibinfo {volume} {47}},\ \bibinfo {pages} {424019}
  (\bibinfo {year} {2014})}\BibitemShut {NoStop}%
\bibitem [{\citenamefont {Gruca}\ \emph {et~al.}(2012)\citenamefont {Gruca},
  \citenamefont {Laskowski},\ and\ \citenamefont {\.Zukowski}}]{Gruca12}%
  \BibitemOpen
  \bibfield  {author} {\bibinfo {author} {\bibfnamefont {J.}~\bibnamefont
  {Gruca}}, \bibinfo {author} {\bibfnamefont {W.}~\bibnamefont {Laskowski}}, \
  and\ \bibinfo {author} {\bibfnamefont {M.}~\bibnamefont {\.Zukowski}},\
  }\href {\doibase 10.1103/PhysRevA.85.022118} {\bibfield  {journal} {\bibinfo
  {journal} {Phys. Rev. A}\ }\textbf {\bibinfo {volume} {85}},\ \bibinfo
  {pages} {022118} (\bibinfo {year} {2012})}\BibitemShut {NoStop}%
\bibitem [{\citenamefont {Masanes}\ \emph {et~al.}(2008)\citenamefont
  {Masanes}, \citenamefont {Liang},\ and\ \citenamefont {Doherty}}]{Masanes08}%
  \BibitemOpen
  \bibfield  {author} {\bibinfo {author} {\bibfnamefont {L.}~\bibnamefont
  {Masanes}}, \bibinfo {author} {\bibfnamefont {Y.-C.}\ \bibnamefont {Liang}},
  \ and\ \bibinfo {author} {\bibfnamefont {A.~C.}\ \bibnamefont {Doherty}},\
  }\href {\doibase 10.1103/PhysRevLett.100.090403} {\bibfield  {journal}
  {\bibinfo  {journal} {Phys. Rev. Lett.}\ }\textbf {\bibinfo {volume} {100}},\
  \bibinfo {pages} {090403} (\bibinfo {year} {2008})}\BibitemShut {NoStop}%
\bibitem [{\citenamefont {Fonseca}\ and\ \citenamefont
  {Parisio}(2015)}]{CGLMPvolume}%
  \BibitemOpen
  \bibfield  {author} {\bibinfo {author} {\bibfnamefont {E.~A.}\ \bibnamefont
  {Fonseca}}\ and\ \bibinfo {author} {\bibfnamefont {F.}~\bibnamefont
  {Parisio}},\ }\href {\doibase 10.1103/PhysRevA.92.030101} {\bibfield
  {journal} {\bibinfo  {journal} {Phys. Rev. A}\ }\textbf {\bibinfo {volume}
  {92}},\ \bibinfo {pages} {030101} (\bibinfo {year} {2015})}\BibitemShut
  {NoStop}%
\bibitem [{\citenamefont {Kaszlikowski}\ \emph {et~al.}(2000)\citenamefont
  {Kaszlikowski}, \citenamefont {Gnaci\ifmmode~\acute{n}\else \'{n}\fi{}ski},
  \citenamefont {\ifmmode~\dot{Z}\else \.{Z}\fi{}ukowski}, \citenamefont
  {Miklaszewski},\ and\ \citenamefont {Zeilinger}}]{Kaszlikowski00}%
  \BibitemOpen
  \bibfield  {author} {\bibinfo {author} {\bibfnamefont {D.}~\bibnamefont
  {Kaszlikowski}}, \bibinfo {author} {\bibfnamefont {P.}~\bibnamefont
  {Gnaci\ifmmode~\acute{n}\else \'{n}\fi{}ski}}, \bibinfo {author}
  {\bibfnamefont {M.}~\bibnamefont {\ifmmode~\dot{Z}\else \.{Z}\fi{}ukowski}},
  \bibinfo {author} {\bibfnamefont {W.}~\bibnamefont {Miklaszewski}}, \ and\
  \bibinfo {author} {\bibfnamefont {A.}~\bibnamefont {Zeilinger}},\ }\href
  {\doibase 10.1103/PhysRevLett.85.4418} {\bibfield  {journal} {\bibinfo
  {journal} {Phys. Rev. Lett.}\ }\textbf {\bibinfo {volume} {85}},\ \bibinfo
  {pages} {4418} (\bibinfo {year} {2000})}\BibitemShut {NoStop}%
\bibitem [{\citenamefont {Masanes}(2002)}]{Masanes02}%
  \BibitemOpen
  \bibfield  {author} {\bibinfo {author} {\bibfnamefont {L.}~\bibnamefont
  {Masanes}},\ }\href@noop {} {\bibfield  {journal} {\bibinfo  {journal}
  {Quant. Inf. Comp.}\ }\textbf {\bibinfo {volume} {3}},\ \bibinfo {pages}
  {345} (\bibinfo {year} {2002})}\BibitemShut {NoStop}%
\bibitem [{\citenamefont {Ac\'{\i}n}\ \emph {et~al.}(2002)\citenamefont
  {Ac\'{\i}n}, \citenamefont {Durt}, \citenamefont {Gisin},\ and\ \citenamefont
  {Latorre}}]{CGLMPacin}%
  \BibitemOpen
  \bibfield  {author} {\bibinfo {author} {\bibfnamefont {A.}~\bibnamefont
  {Ac\'{\i}n}}, \bibinfo {author} {\bibfnamefont {T.}~\bibnamefont {Durt}},
  \bibinfo {author} {\bibfnamefont {N.}~\bibnamefont {Gisin}}, \ and\ \bibinfo
  {author} {\bibfnamefont {J.~I.}\ \bibnamefont {Latorre}},\ }\href {\doibase
  10.1103/PhysRevA.65.052325} {\bibfield  {journal} {\bibinfo  {journal} {Phys.
  Rev. A}\ }\textbf {\bibinfo {volume} {65}},\ \bibinfo {pages} {052325}
  (\bibinfo {year} {2002})}\BibitemShut {NoStop}%
\bibitem [{\citenamefont {Hu}\ \emph {et~al.}(2008)\citenamefont {Hu},
  \citenamefont {Deng},\ and\ \citenamefont {Chen}}]{CGLMPmax08}%
  \BibitemOpen
  \bibfield  {author} {\bibinfo {author} {\bibfnamefont {M.-G.}\ \bibnamefont
  {Hu}}, \bibinfo {author} {\bibfnamefont {D.-L.}\ \bibnamefont {Deng}}, \ and\
  \bibinfo {author} {\bibfnamefont {J.-L.}\ \bibnamefont {Chen}},\ }\href@noop
  {} {\bibfield  {journal} {\bibinfo  {journal} {Int. J. Quant. Inf.}\ }\textbf
  {\bibinfo {volume} {6}},\ \bibinfo {pages} {1067} (\bibinfo {year}
  {2008})}\BibitemShut {NoStop}%
\bibitem [{\citenamefont {Spengler}\ \emph {et~al.}(2011)\citenamefont
  {Spengler}, \citenamefont {Huber},\ and\ \citenamefont
  {Hiesmayr}}]{Spengler11}%
  \BibitemOpen
  \bibfield  {author} {\bibinfo {author} {\bibfnamefont {C.}~\bibnamefont
  {Spengler}}, \bibinfo {author} {\bibfnamefont {M.}~\bibnamefont {Huber}}, \
  and\ \bibinfo {author} {\bibfnamefont {B.~C.}\ \bibnamefont {Hiesmayr}},\
  }\href {\doibase 10.1088/1751-8113/44/6/065304} {\bibfield  {journal}
  {\bibinfo  {journal} {J. Phys. A: Math. Theor.}\ }\textbf {\bibinfo {volume}
  {44}},\ \bibinfo {pages} {065304} (\bibinfo {year} {2011})}\BibitemShut
  {NoStop}%
\bibitem [{\citenamefont {Braunstein}\ \emph {et~al.}(1992)\citenamefont
  {Braunstein}, \citenamefont {Mann},\ and\ \citenamefont
  {Revzen}}]{Braunstein92}%
  \BibitemOpen
  \bibfield  {author} {\bibinfo {author} {\bibfnamefont {S.~L.}\ \bibnamefont
  {Braunstein}}, \bibinfo {author} {\bibfnamefont {A.}~\bibnamefont {Mann}}, \
  and\ \bibinfo {author} {\bibfnamefont {M.}~\bibnamefont {Revzen}},\ }\href
  {\doibase 10.1103/PhysRevLett.68.3259} {\bibfield  {journal} {\bibinfo
  {journal} {Phys. Rev. Lett.}\ }\textbf {\bibinfo {volume} {68}},\ \bibinfo
  {pages} {3259} (\bibinfo {year} {1992})}\BibitemShut {NoStop}%
\bibitem [{\citenamefont {Kitagawa}\ and\ \citenamefont
  {Ueda}(1993)}]{SpinSqueezed93}%
  \BibitemOpen
  \bibfield  {author} {\bibinfo {author} {\bibfnamefont {M.}~\bibnamefont
  {Kitagawa}}\ and\ \bibinfo {author} {\bibfnamefont {M.}~\bibnamefont
  {Ueda}},\ }\href {\doibase 10.1103/PhysRevA.47.5138} {\bibfield  {journal}
  {\bibinfo  {journal} {Phys. Rev. A}\ }\textbf {\bibinfo {volume} {47}},\
  \bibinfo {pages} {5138} (\bibinfo {year} {1993})}\BibitemShut {NoStop}%
\bibitem [{\citenamefont {Kurzy\ifmmode~\acute{n}\else \'{n}\fi{}ski}\ \emph
  {et~al.}(2016)\citenamefont {Kurzy\ifmmode~\acute{n}\else \'{n}\fi{}ski},
  \citenamefont {Ko\l{}odziejski}, \citenamefont {Laskowski},\ and\
  \citenamefont {Markiewicz}}]{Kurzynski16}%
  \BibitemOpen
  \bibfield  {author} {\bibinfo {author} {\bibfnamefont {P.}~\bibnamefont
  {Kurzy\ifmmode~\acute{n}\else \'{n}\fi{}ski}}, \bibinfo {author}
  {\bibfnamefont {A.}~\bibnamefont {Ko\l{}odziejski}}, \bibinfo {author}
  {\bibfnamefont {W.}~\bibnamefont {Laskowski}}, \ and\ \bibinfo {author}
  {\bibfnamefont {M.}~\bibnamefont {Markiewicz}},\ }\href {\doibase
  10.1103/PhysRevA.93.062126} {\bibfield  {journal} {\bibinfo  {journal} {Phys.
  Rev. A}\ }\textbf {\bibinfo {volume} {93}},\ \bibinfo {pages} {062126}
  (\bibinfo {year} {2016})}\BibitemShut {NoStop}%
\bibitem [{\citenamefont {Clauser}\ \emph {et~al.}(1969)\citenamefont
  {Clauser}, \citenamefont {Horne}, \citenamefont {Shimony},\ and\
  \citenamefont {Holt}}]{CHSH}%
  \BibitemOpen
  \bibfield  {author} {\bibinfo {author} {\bibfnamefont {J.~F.}\ \bibnamefont
  {Clauser}}, \bibinfo {author} {\bibfnamefont {M.~A.}\ \bibnamefont {Horne}},
  \bibinfo {author} {\bibfnamefont {A.}~\bibnamefont {Shimony}}, \ and\
  \bibinfo {author} {\bibfnamefont {R.~A.}\ \bibnamefont {Holt}},\ }\href
  {\doibase 10.1103/PhysRevLett.23.880} {\bibfield  {journal} {\bibinfo
  {journal} {Phys. Rev. Lett.}\ }\textbf {\bibinfo {volume} {23}},\ \bibinfo
  {pages} {880} (\bibinfo {year} {1969})}\BibitemShut {NoStop}%
\bibitem [{\citenamefont {Mermin}(1990)}]{Mermin90}%
  \BibitemOpen
  \bibfield  {author} {\bibinfo {author} {\bibfnamefont {N.~D.}\ \bibnamefont
  {Mermin}},\ }\href {\doibase 10.1103/PhysRevLett.65.1838} {\bibfield
  {journal} {\bibinfo  {journal} {Phys. Rev. Lett.}\ }\textbf {\bibinfo
  {volume} {65}},\ \bibinfo {pages} {1838} (\bibinfo {year}
  {1990})}\BibitemShut {NoStop}%
\bibitem [{\citenamefont {Kurzy\ifmmode~\acute{n}\else \'{n}\fi{}ski}\ \emph
  {et~al.}(2011)\citenamefont {Kurzy\ifmmode~\acute{n}\else \'{n}\fi{}ski},
  \citenamefont {Paterek}, \citenamefont {Ramanathan}, \citenamefont
  {Laskowski},\ and\ \citenamefont {Kaszlikowski}}]{BellComplementarity}%
  \BibitemOpen
  \bibfield  {author} {\bibinfo {author} {\bibfnamefont {P.}~\bibnamefont
  {Kurzy\ifmmode~\acute{n}\else \'{n}\fi{}ski}}, \bibinfo {author}
  {\bibfnamefont {T.}~\bibnamefont {Paterek}}, \bibinfo {author} {\bibfnamefont
  {R.}~\bibnamefont {Ramanathan}}, \bibinfo {author} {\bibfnamefont
  {W.}~\bibnamefont {Laskowski}}, \ and\ \bibinfo {author} {\bibfnamefont
  {D.}~\bibnamefont {Kaszlikowski}},\ }\href {\doibase
  10.1103/PhysRevLett.106.180402} {\bibfield  {journal} {\bibinfo  {journal}
  {Phys. Rev. Lett.}\ }\textbf {\bibinfo {volume} {106}},\ \bibinfo {pages}
  {180402} (\bibinfo {year} {2011})}\BibitemShut {NoStop}%
\bibitem [{\citenamefont {Bertlmann}\ and\ \citenamefont
  {Krammer}(2008)}]{Krammer08}%
  \BibitemOpen
  \bibfield  {author} {\bibinfo {author} {\bibfnamefont {R.~A.}\ \bibnamefont
  {Bertlmann}}\ and\ \bibinfo {author} {\bibfnamefont {P.}~\bibnamefont
  {Krammer}},\ }\href@noop {} {\bibfield  {journal} {\bibinfo  {journal} {J.
  Phys. A: Math. Theor.}\ }\textbf {\bibinfo {volume} {41}},\ \bibinfo {pages}
  {235303} (\bibinfo {year} {2008})}\BibitemShut {NoStop}%
\bibitem [{\citenamefont {Laskowski}\ \emph {et~al.}(2013)\citenamefont
  {Laskowski}, \citenamefont {Markiewicz}, \citenamefont {Paterek},\ and\
  \citenamefont {Weinar}}]{LM13}%
  \BibitemOpen
  \bibfield  {author} {\bibinfo {author} {\bibfnamefont {W.}~\bibnamefont
  {Laskowski}}, \bibinfo {author} {\bibfnamefont {M.}~\bibnamefont
  {Markiewicz}}, \bibinfo {author} {\bibfnamefont {T.}~\bibnamefont {Paterek}},
  \ and\ \bibinfo {author} {\bibfnamefont {R.}~\bibnamefont {Weinar}},\ }\href
  {\doibase 10.1103/PhysRevA.88.022304} {\bibfield  {journal} {\bibinfo
  {journal} {Phys. Rev. A}\ }\textbf {\bibinfo {volume} {88}},\ \bibinfo
  {pages} {022304} (\bibinfo {year} {2013})}\BibitemShut {NoStop}%
\bibitem [{\citenamefont {Chen}\ \emph {et~al.}(2006)\citenamefont {Chen},
  \citenamefont {Wu}, \citenamefont {Kwek}, \citenamefont {Oh},\ and\
  \citenamefont {Ge}}]{Chen06}%
  \BibitemOpen
  \bibfield  {author} {\bibinfo {author} {\bibfnamefont {J.-L.}\ \bibnamefont
  {Chen}}, \bibinfo {author} {\bibfnamefont {C.}~\bibnamefont {Wu}}, \bibinfo
  {author} {\bibfnamefont {L.~C.}\ \bibnamefont {Kwek}}, \bibinfo {author}
  {\bibfnamefont {C.~H.}\ \bibnamefont {Oh}}, \ and\ \bibinfo {author}
  {\bibfnamefont {M.-L.}\ \bibnamefont {Ge}},\ }\href {\doibase
  10.1103/PhysRevA.74.032106} {\bibfield  {journal} {\bibinfo  {journal} {Phys.
  Rev. A}\ }\textbf {\bibinfo {volume} {74}},\ \bibinfo {pages} {032106}
  (\bibinfo {year} {2006})}\BibitemShut {NoStop}%
\bibitem [{\citenamefont {Toner}\ and\ \citenamefont
  {Verstraete}(2006)}]{Toner06}%
  \BibitemOpen
  \bibfield  {author} {\bibinfo {author} {\bibfnamefont {B.}~\bibnamefont
  {Toner}}\ and\ \bibinfo {author} {\bibfnamefont {F.}~\bibnamefont
  {Verstraete}},\ }\href@noop {} {\bibfield  {journal} {\bibinfo  {journal}
  {arXiv:quant-ph/0611001}\ } (\bibinfo {year} {2006})}\BibitemShut {NoStop}%
\bibitem [{\citenamefont {Chen}\ \emph {et~al.}(2008)\citenamefont {Chen},
  \citenamefont {Wu}, \citenamefont {Kwek},\ and\ \citenamefont {Oh}}]{Chen08}%
  \BibitemOpen
  \bibfield  {author} {\bibinfo {author} {\bibfnamefont {J.-L.}\ \bibnamefont
  {Chen}}, \bibinfo {author} {\bibfnamefont {C.}~\bibnamefont {Wu}}, \bibinfo
  {author} {\bibfnamefont {L.~C.}\ \bibnamefont {Kwek}}, \ and\ \bibinfo
  {author} {\bibfnamefont {C.~H.}\ \bibnamefont {Oh}},\ }\href {\doibase
  10.1103/PhysRevA.78.032107} {\bibfield  {journal} {\bibinfo  {journal} {Phys.
  Rev. A}\ }\textbf {\bibinfo {volume} {78}},\ \bibinfo {pages} {032107}
  (\bibinfo {year} {2008})}\BibitemShut {NoStop}%
\end{thebibliography}
	
\end{document}